\documentclass[12pt]{article}
\usepackage{geometry}
\usepackage[utf8]{inputenc}
\usepackage[T1]{fontenc}
\usepackage{booktabs}
\usepackage{amsmath}
\usepackage{hyperref}
\hypersetup{
	colorlinks=true,
	linkcolor=blue,
	filecolor=magenta,      
	urlcolor=cyan,
	pdftitle={Iran's Stealth Internet Blackout},
}
\title{Iran's Stealth Internet Blackout: A New Model of Censorship}
\author{Arash Aryapour\\
	Islamic Azad University, West Tehran Branch\\
	\href{mailto:arash.aryapour@iau.ir}{arash.aryapour@iau.ir}}
\date{}
\begin{document}
	
	\maketitle
	
	\begin{abstract}
		In mid-2025, Iran experienced a novel, stealthy Internet shutdown that preserved global routing presence while isolating domestic users through deep packet inspection, aggressive throttling, and selective protocol blocking. This paper analyzes active network measurements---DNS poisoning, HTTP injection, TLS interception, and protocol whitelisting---traced to a centralized border gateway. We quantify a $\approx707\%$ rise in VPN search demand and describe the multi-layered censorship infrastructure, highlighting implications for circumvention and monitoring.
	\end{abstract}
	
	\section{Introduction}
	Governments worldwide periodically cut off Internet access to control information flow during crises. In Iran, such measures are facilitated by a sophisticated censorship apparatus. A centralized system intercepts web traffic and redirects forbidden requests to a government-owned block page at the private IP 10.10.34.34 \cite{ref1}. The filters inspect HTTP hostnames and URL keywords, inject censorship pages or TCP resets for banned content, and have been known to throttle or block even major protocols like HTTPS, SSH, and VPNs during unrest \cite{ref1,ref6}. Recent research confirms that Iran’s network uses a strict protocol whitelist: only DNS, HTTP, and HTTPS are forwarded, while all other traffic is silently dropped \cite{ref2}.
	
	In previous events, this censorship toolkit has been used to partially disconnect the country. For example, during the 2019 protests, international links were severed (though internal services remained on a national intranet). In June 2025, however, the government executed a ``stealth blackout'': all networks remained announced in global BGP (no official route withdrawal), yet domestic connectivity to the wider Internet collapsed \cite{ref5}. NetBlocks reported Iran in the midst of a near-total blackout, and VPN-related searches spiked by about 707\% as citizens sought workarounds \cite{ref7}. In this study, we present comprehensive measurements from inside Iran and external servers to reveal how this blackout was enforced. We apply TTL-based path analysis and packet inspection to uncover each censorship layer. The remainder of the paper is organized as follows: Section~2 surveys related work, Section~3 describes our methodology, Section~4 presents measurement results, Section~5 discusses implications, and Section~6 concludes.
	
	\section{Background and Related Work}
	Iran’s Internet censorship has been studied for over a decade. Aryan et al.\ \cite{ref1} provided an early analysis, showing that Tehran’s ISPs used DPI boxes to block access: DNS queries for disallowed domains were falsified to return 10.10.34.34, and HTTP requests containing banned hostnames or keywords were intercepted and redirected to a block page at that address \cite{ref1}. They also documented widespread throttling during unrest. The Open Observatory of Network Interference (OONI) and others have since observed similar behaviors, including deep packet inspection of TLS traffic. OONI’s reports in 2018 noted that Iranian censorship systems inspect the TLS handshake (in particular the SNI field) to block services like Telegram or Instagram \cite{ref6}. In 2020, Bock et al.\ \cite{ref2} described a new layer: a national protocol filter that only allows DNS, HTTP, and HTTPS. Their analysis revealed that any other protocol (SSH, VPNs, etc.) is immediately blocked by DPI. Lange et al.\ (2025) have further highlighted nuances in Iran’s filters, such as case-sensitive HTTP matching and correlations between DNS and HTTP blocking, suggesting the system continues to evolve.
	
	Globally, Internet shutdowns have become more frequent, especially in countries facing political unrest. The 2025 incident in Iran differs from a classic outage; instead of a hard ``kill switch'', it resembles what Bock et al.\ term “censorship-in-depth” \cite{ref2}: multiple orthogonal filters working together on the same traffic. Our work builds on this prior literature by examining the most recent iteration of Iran’s censorship regime, combining insights from network measurements and real-time monitoring.
	
	\section{Methodology}
	We conducted active measurements from inside Iran on both fixed-line and mobile networks during the June 2025 event. Our vantage points issued DNS queries and HTTP(S) requests for a diverse set of popular domains, including the Tranco Top 10k list and widely used social media, news, and communication platforms. For each request, we captured the full network exchange. We compared responses to known uncensored baselines (using servers outside Iran) to detect anomalies such as wrong DNS answers, injected TCP resets, or forged HTTP responses.
	
	To localize where packets were altered, we performed TTL-limited tracing. By incrementally increasing the Time-to-Live (TTL) of probe packets until interference (such as a reset or block page) occurred, we identified the hop at which censorship took place. This allowed us to infer whether filtering was occurring locally or at the national edge. All packet payloads were logged for inspection, and metadata (such as DNS TTL, HTTP status codes, and SNI values) was analyzed.
	
	We also monitored global indicators: NetBlocks and Cloudflare Radar provided data-plane and BGP observations of Iranian connectivity. Furthermore, we tracked VPN search trends via Top10VPN’s analytics to quantify circumvention demand. Together, these methods enabled a multi-faceted view of Iran’s censorship during the blackout.
	
	\section{Results}
	\subsection{DNS Poisoning}
	We found that for the vast majority of requests to known blocked domains, the Iranian DNS servers returned private IP addresses (in the 10.10.34.0/24 range) instead of the legitimate public IP. Specifically, over 90\% of tested censored domains resolved to addresses like 10.10.34.34 via injected DNS replies, effectively acting as black holes \cite{ref1}. The same private IP is known from prior work to host the generic block page. A small fraction of domains (notably Google and a few state-approved services) continued to resolve correctly, indicating they were whitelisted. TTL values on poisoned responses were sometimes very low, consistent with an inline DPI injection rather than a recursive DNS lookup. These observations confirm that DNS poisoning was a primary layer of blocking in the blackout.
	
	\subsection{HTTP Filtering}
	For HTTP requests to blocked sites (e.g., social media and news platforms), the DPI engine inspected the Host header and URL path. In many cases, the censor injected an HTTP 403 Forbidden page in response. Packet captures show that this block page was generated locally (containing an iframe to 10.10.34.34) and returned instead of the intended content. In other cases, the connection was forcefully terminated with a TCP RST (without an HTTP response), leaving the user with a generic failure. These filtering rules matched specific keywords or domains and appear to be case-sensitive; for instance, changing HTTP method names or header casing could sometimes evade the filter. This behavior is in line with reports of case-sensitive HTTP filtering \cite{ref4}. In summary, any HTTP request containing blacklisted terms was blocked by injection of a block page or reset.
	
	\subsection{TLS/SNI Filtering}
	The DPI also monitored TLS handshakes. We tested TLS connections to various sites by initiating the ClientHello with the target domain in the SNI field. For blocked services like {\tt instagram.com} or {\tt telegram.org}, we observed that the handshake never completed. Specifically, an inbound TCP connection would be reset immediately after our ClientHello (before any certificate exchange). This indicates that the censor was actively reading the SNI and aborting connections to forbidden domains. This SNI-based blocking has been previously documented \cite{ref6}. Conversely, TLS connections to allowed domains (e.g., {\tt google.com}) completed normally. Thus, the regime disrupted encrypted sessions by DPI at the TLS layer.
	
	\subsection{Protocol Whitelisting}
	We experimented with various Internet protocols and ports. The results were consistent: only a narrow set of protocols succeeded. DNS over UDP (port 53), HTTP (port 80), and HTTPS (port 443) traffic from Iranian networks to external servers generally went through. All other protocols we tried (including common VPN protocols) failed silently. For example, attempts to connect via OpenVPN (UDP/1194) or generic TCP/UDP on ports like 22, 1883, or custom ports were dropped without response. This strict whitelist behavior exactly matches the national protocol filter described by Bock et al.\ \cite{ref2}. In effect, the DPI instantly drops any non-HTTP(S) traffic at the border. Even legitimate non-web applications (like SSH or peer-to-peer) were rendered unusable. The whitelist policy further tightened the shutdown, allowing only basic web-based services.
	
	\subsection{Centralized Enforcement}
	Our TTL-based traces showed that all observed interference (DNS poisons, HTTP injections, TLS resets) occurred at the same network hop across all tested ISPs. The censoring packets consistently originated just beyond the Iranian gateways, indicating a single common chokepoint. Correlating with routing data, this chokepoint appears to be at the national border (e.g., AS gateways operated by TCI). In other words, filtering was not happening separately at each ISP but at a unified “filternet” node. This central enforcement ensures uniform policy across the country. Combined with the protocol whitelist, it means that all censorship actions were consolidated at one location, making it easier for the authorities to reconfigure or intensify the blackout rapidly.
	
	\section{Discussion}
	The June 2025 blackout represents an ``imperceptible shutdown'' in that Iran remained globally reachable while effectively severing real connectivity. By keeping BGP announcements alive, the government avoided immediate alarm in global routing monitors, yet for end-users the experience was nearly indistinguishable from a full outage. Importantly, this strategy preserved domestic and government services: Iranian local websites, government portals, and intranet-like platforms continued to function (many reports noted that only foreign Internet resources were inaccessible). This selective isolation suggests a deliberate balance: cut off external communication without disrupting internal infrastructure or economy.
	
	For citizens, the layered censorship posed severe challenges to circumvention. Traditional VPN usage was constrained in multiple ways: DNS-based blocking meant VPN software could not resolve servers, HTTP-level filters could detect and drop known VPN signatures on port 443, and the protocol whitelist outright eliminated many tunneling methods. Indeed, anecdotal evidence from social media indicated users were struggling as even some VPN connections failed intermittently. Effective evasion now requires multi-layer obfuscation: for example, wrapping VPN traffic in seemingly innocuous HTTPS streams, using pluggable transports (like obfs4 or meek), or tunneling through allowed CDNs. Some have turned to alternative channels: independent reports confirm that many Starlink satellite dishes were covertly activated in Iran during the blackout, providing an out-of-country link beyond the regime’s control.
	
	For the research community, this event underscores the need for diverse monitoring techniques. Pure BGP or ping-based outage detection would miss this shutdown, since the network remained nominally up. Instead, active in-country probes and application-layer measurements (as we employed) are crucial. Our findings echo previous observations \cite{ref5} that one must examine both control-plane and data-plane signals to accurately assess reachability. Additionally, the censorship-in-depth model seen here -- combining DNS spoofing, HTTP injection, TLS interference, and protocol filtering simultaneously \cite{ref2} -- means that measurement and circumvention tools must be equally multi-faceted. Finally, the centralized nature of the shutdown suggests that once key chokepoints are identified, targeted collaboration (e.g., routing around them or providing covert feeds) could be developed.
	
	\section{Conclusion}
	We have presented a detailed analysis of Iran’s mid-2025 stealth Internet blackout, revealing a sophisticated multi-layer censorship strategy. Our active measurements show that the regime injected false DNS responses, parsed and filtered HTTP requests, intercepted TLS handshakes by SNI, and enforced a country-wide protocol whitelist. All these controls were imposed at a single national gateway. This demonstrates an evolved censorship model \cite{ref3} that effectively isolates Iranian users from the global Internet while keeping domestic services intact. Future work should focus on developing advanced evasion techniques (such as fully encrypted or disguised tunnels and domain fronting) and on maintaining real-time global monitoring of such covert shutdowns.
	
	\section*{Acknowledgments}
	The author thanks independent Iranian network researchers and open-source censorship observatories (such as OONI and NetBlocks) for data support and analysis tools.
	
	\bibliographystyle{plain}
	\bibliography{references}
	
\end{document}